*SiPM and PMT Driving, Signals Count, and Peak Detection Circuits, suitable for Particle Detection.*

**Gholamreza Fardipour Raki[1], Maryam Ghahremani Gol, Mohammad Sahraei, Mohsen Khakzad**

*School of particles and accelerator, Institute for Research in fundamental sciences (IPM), P.O.Box 19395-5531, Tehran, Iran*
Email: fardipour@ipm.ir

**Abstract**: The signals received from the optical receivers like SiPM and PMT due to the collision of energetic particles with the scintillators attached to these optical receivers are weak and fast. Optimizing signals is necessary to measure the number of signals and their peak height with electronic circuits. This text presents an example of SiPM's driver circuit, signal counting, and peak measurement. Also, the electronic circuits necessary to optimize the signals, including amplification, removing background noise, converting the signal to digital, and increasing the duration of the signal, are presented in this text. In the end, we provide two tests to confirm the correct operation of the circuits. Such a system has several advantages. This set has a small volume and is portable. Its operating voltage is 12 volts, with a current of about 0.3 amps; As a result, it is easily possible to use this set in any experiment. In addition, the cost of building such a system is much lower than providing similar ready-made designs. The most important achievement here is to convert the standard signal taken from the detector into an almost ideal optimized signal for signal counting and peak measurement. Therefore, it seems that using all or part of these circuits can be helpful for researchers. This text presents a particular method for signal optimization and provides the reader with a coherent and complete process of building and testing circuits. If the reader is familiar with the basics of electronics and detectors, they can reconstruct the circuits without any problems. Therefore, parts of this text may have an educational and review form.

---

[1] . Corresponding Author

# Contents:

## 1. PMT and SiPM signals

PMT (Photomultiplier tube) and SiPM (Silicon Photomultiplier) can detect even single photons. In photon detection experiments with these instruments, the number of photons is sometimes small. The time of photon generation and absorption by the medium is very short, so the signals obtained from PMT and SiPM are in the range of 100 mV peak and 100 ns duration, but the voltage and time interval of the signals may not be regular. Usually, there is not much regularity in the form of signals that come from the detection of energetic particles such as cosmic muons or particles from the decay of radioactive sources with plastic scintillators because of fluctuations of energy loss of charged particles in matter. The number of photons produced in the scintillator is proportional to scintillator thickness, particle path, and energy [1].The number of photons arriving at the PMT or SiPM almost simultaneously determines the maximum voltage and duration of the signal. The signals can have a negative or positive peak depending on the structure of the photomultiplier driver circuit. PMTs commonly generate negative signals [2].Due to the high application of these components rather than SiPM, here we designed the SiPM driver circuits to generate PMT-like signals so that for subsequent processes and signal analysis, both types of photomultipliers can be used in the same way.

Usually, for each type of PMT, different driver circuits are introduced by the manufacturer. Applications considered on the measurement of particle energy (equivalent to the number of photons reaching the optical receiver) have a different circuit from applications based on signal counting. Generating signals with low noise and ring and higher peak voltage is appropriate for signal counting. However, in energy measurement, the signal peak must be as linear as possible to the number of photons reaching the PMT [2]. Therefore, using the circuits recommended by the manufacturer to design the PMT driver circuit seems appropriate according to the type of application. In addition, the bias voltage is very effective for adjusting the PMT's peak signal height and noise level. Using the maximum tolerable voltage in the PMT does not necessarily produce a better signal. However, the correct voltage is the maximum bias voltage that produces an acceptable signal shape.

The signals emitted from photomultipliers can be easily affected. One should use high-quality connectors and cables to prevent signal deformation during their initial transmission. These signals are so weak and vulnerable to transmission to subsequent devices. Therefore, in transmitting the signal to the next devices, it is better to use systems with very high input resistance (in the range of 50 to 100 MΩ) at the input of the following devices. This high resistance does not allow significant current to flow, thus minimizes voltage drop and deforms the signal. Nevertheless, in this case, if the next device is an amplifier, the signal amplification cannot be done enough without adding feedback in the amplifier entrance. This feedback will create less input resistance at the entrance of the amplifier.

SiPM driver circuits are completely different from PMT. Usually, the manufacturer of SiPM does not provide the methods to supply the bias voltage, the signal preamplification, and how to use SiPM in a detector. Although they explain how to use the component, in principle, this explanation is not enough to build a driver circuit. Therefore we introduced a particular type of SiPM driver here.

## 2. SiPM Detector Circuit

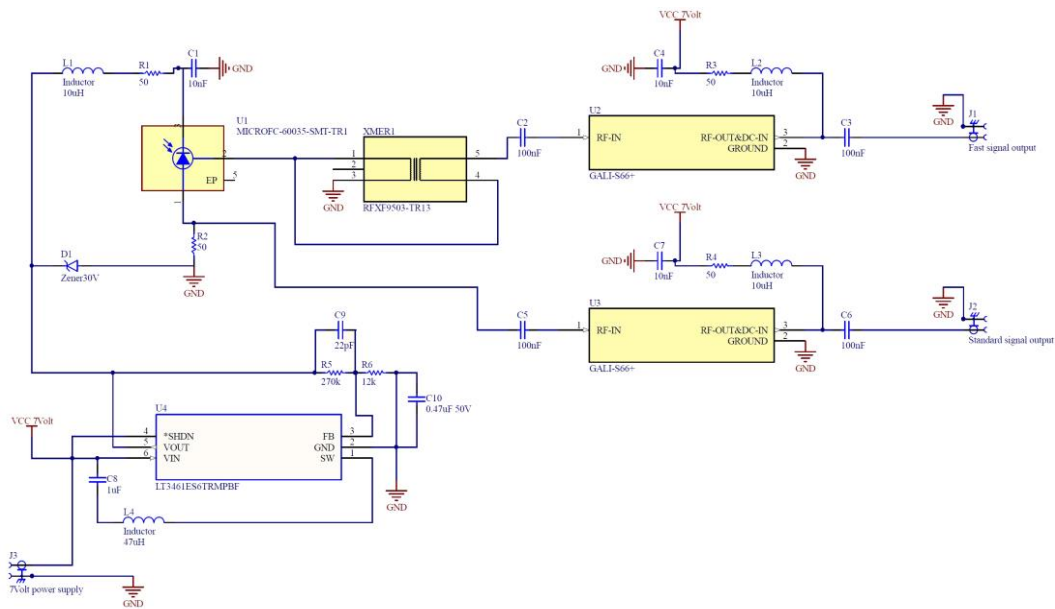

**Figure 1.** The complete circuit for driving SiPM, contains biasing voltage and preamplifiers for both standard and fast outputs.

Here is the driving and preamplifier circuit for SiPM with ON-Semiconductor Micro FC60. Additional information about the structure of SiPM and exclusively this device, along with necessary explanations about its operating conditions, are provided by the manufacturer company [3, 4, 5]. The structure of SiPMs is very similar, and the driver circuit presented here can be used for any type of SiPM with minor modifications. These modifications are only related to the drive voltage or component biasing. SiPM is made of a two-dimensional array of a large number of sensitive photodiodes.

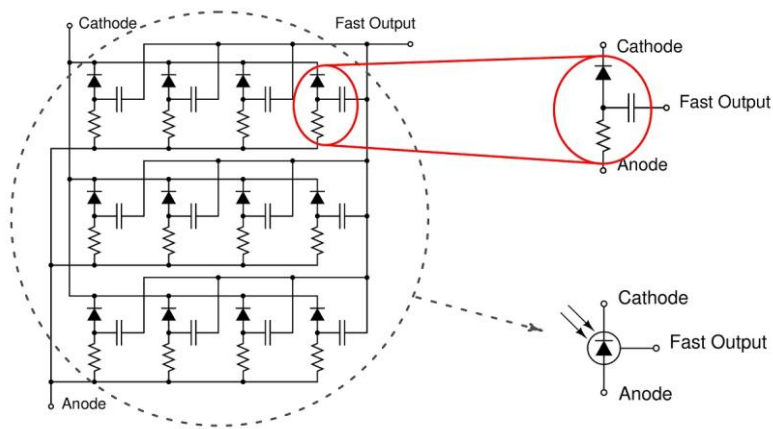

**Figure 2.** SiPM internal structure and component symbol. The SiPM consists of an array of microcells connected in parallel.

The diodes are connected in parallel, and their effect is summed at the output. The greater the number of affected diodes at the same time, the higher the peak voltage of the output signal. Photodiodes can be used in forward or reverse bias. In forward bias, the greater the number of photons in the diode's frequency sensitivity range, the stronger the signal (due to the reduction in the diode's internal resistance). However, in this way, the effect of a small number of photons is never detectable.

The photodiode must be advanced to the breakdown limit in the reverse bias method. This reverse voltage for the SiPM used here is about 30 volts. This value must be adjusted very carefully. Voltages above the range specified in the component datasheet can cause severe damage to the component, and lower voltages can significantly reduce signal strength. This voltage is supplied by the DC-DC Booster circuit, which increases the input voltage to the required level. Avalanche can occur in reverse bias and at the breakdown threshold, even when a photon hits the photodiode.

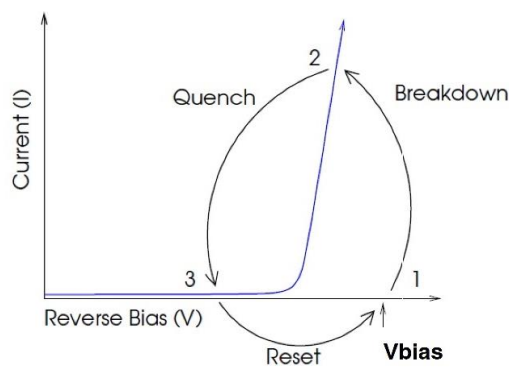

**Figure 3.** Breakdown, Quench and Reset Cycle of a photo diode working in Geiger Mode.

In breakdown, a substantial current compared to the diode's tolerance threshold can pass through the diode and cause damage. However, inside the SiPM, a series resistor is placed for each photodiode, Figure 2 [5]. When the avalanche breakdown starts and the current passes through the diode, the resistance causes the voltage to decrease, according to Ohm's law. This voltage drop takes the diode away from the breakdown limit and stops the current. These processes occur in a SiPM diode in about a few nanoseconds.

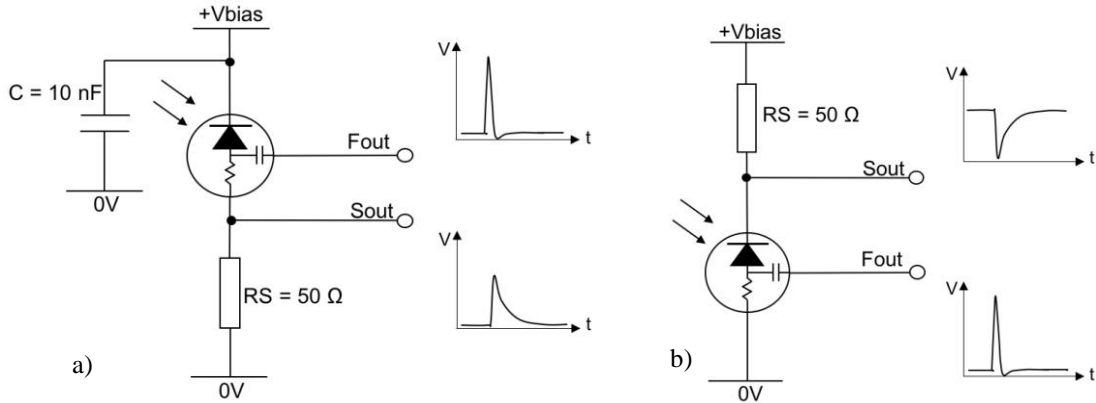

**Figure 4.** Biasing and readout circuit for SiPM. These configuration give the best fast output timing performance.

This process can be performed in one cycle for a diode, repeatedly placed in the ready state for photon detection. This cycle, known as the Geiger mode, is shown in Figure 3 [5] as a voltage-current diagram. SiPMs usually have standard (Sout) and fast (Fout) outputs. The standard output from the anode or cathode junction is shown in Figure 4 [3]. The electrical potential of the junction during an avalanche failure will have stress due to the sudden onset of current. This voltage stress is essentially the same as the signal.

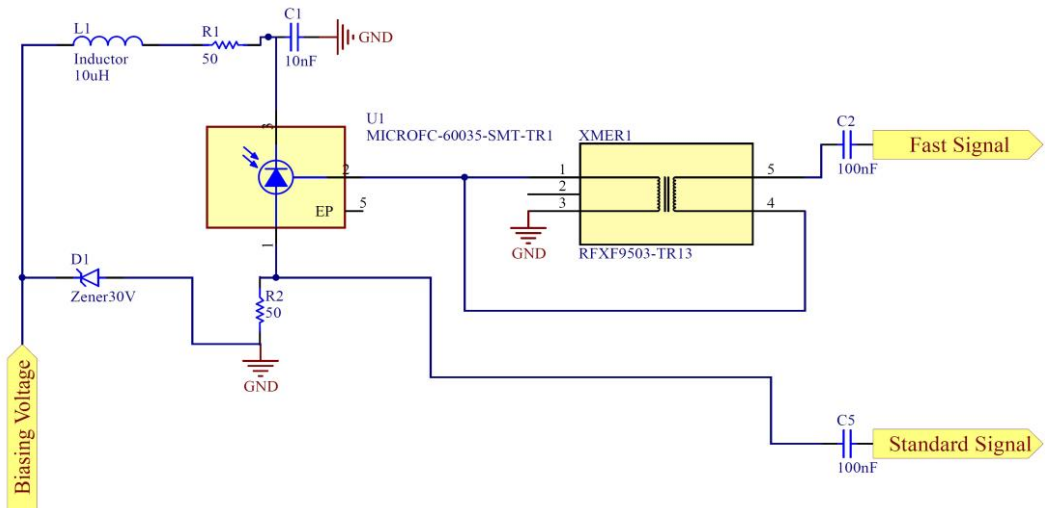

**Figure 5.** Schematic of fabricated biasing and readout circuit for SiPM.

The circuit shown in Figure 5.a has fast and standard positive signals. We used this structure to design a SiPM driver circuit with some further considerations (Figure 5). Transformer XMER1 is recommended for signal transmission over longer distances by the SiPM manufacturer. However, experiments show that this transformer always creates a better signal shape on the fast output. This component is mounted so that the output signal will be positive, but if we replace connections 4 and 5, the signal will be reversed (negative). This component can always be used to reverse fast and weak signals. At the signal outputs, a 100nF capacitor is placed in series. This capacitor is a complete connection for fast signals but simultaneously prevents the actual SiPM connection with subsequent devices. Such conservatism is common in electronics. To build the circuit, we preferred a circular shape with a diameter equivalent to high-consumption PMTs. In this way, we made the boards, including the preamplifiers, the DC-DC Booster, and the connection sockets in layers. As a result, the SiPM detector can be installed precisely instead of the PMT. Figure 6 shows the equivalent circuit of Figure 5.

The power supply of the various electronic circuits introduced here is 3.2, 5, or 7 volts. These voltages are supplied by several small switching power circuits from a 12-volt battery or any equivalent DC source. It is necessary to generate the regulated voltage of 7V by a switching power supply circuit (to ensure that the voltage is stable). This voltage is used to supply the amplifiers and the DC/DC Booster circuit. Thus, it can be said that the operating voltage of the detector is 7 volts.

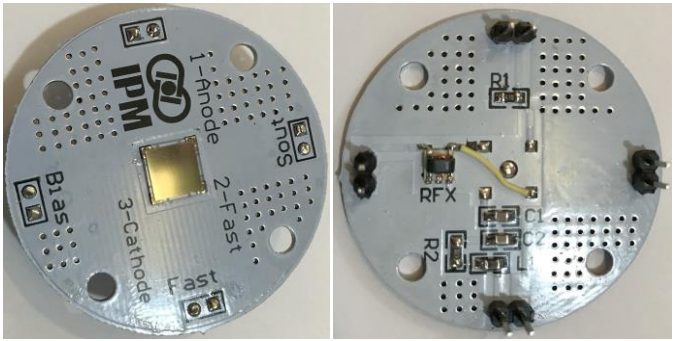

**Figure 6.** Fabricated biasing and readout circuit for SiPM.

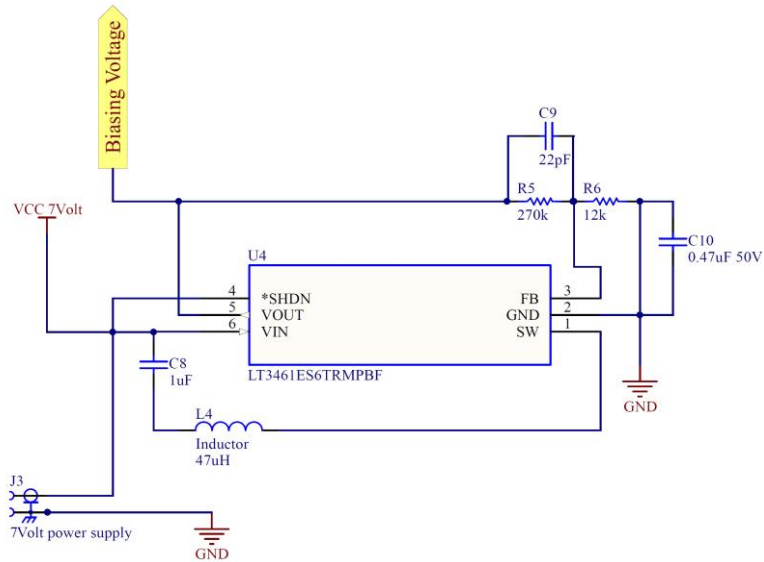

**Figure 7.** Fabricated DC/DC Booster circuit for SiPM.

The booster circuit generates a higher voltage with the same conventional switching method. This voltage can be easily adjusted up to 100 volts. It is only necessary to supply a lower voltage to the Booster IC feedback by dividing the output voltage so that the IC can try to reach the feedback at its nominal voltage by increasing or decreasing the output voltage.

Figure 7 shows the Booster circuit based on the LT3461 IC. Resistors R5 and R6 (voltage divider resistors) should be selected so that the output voltage is 30 volts and the feedback voltage, according to the IC datasheet, is equal to 1.255 volts [6].

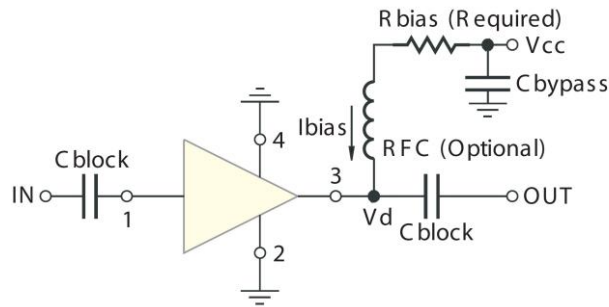

**Figure 8.** GaliS66+ recommended circuit.

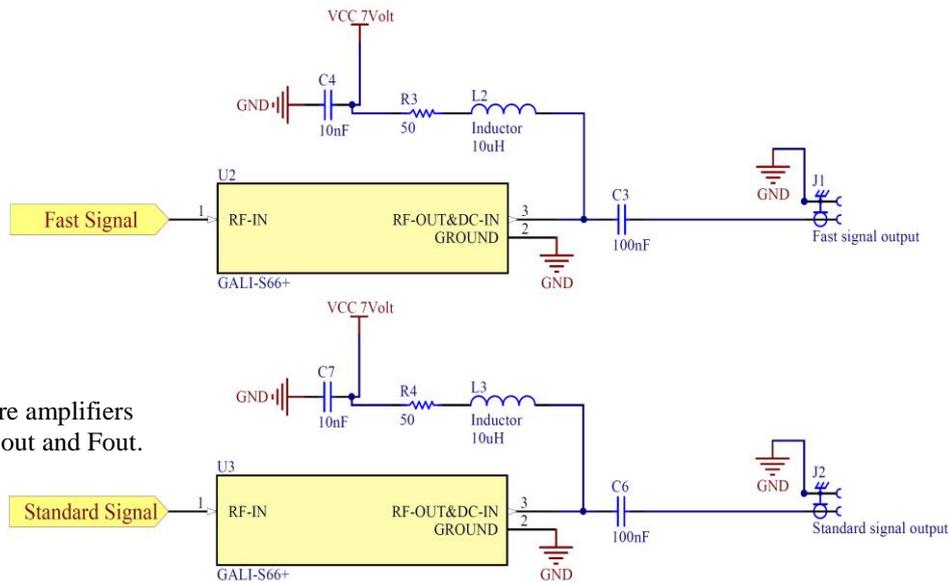

**Figure 9.** Pre amplifiers circuit for Sout and Fout.

Amplifiers usually have a specific operating voltage. The best amplifiers for fast and weak signals are MOSFET-based amplifiers (OpAmp), which will have a minimal effect on the signal shape due to the very high input resistance. Monolithic amplifiers such as Galis66+ have no high input resistance; however, they can amplify short to nanoseconds and long signals up to more than 1 second due to their bandwidth. They also do not have a fixed operating voltage. As shown in Figure 8 [7], the $R_{bias}$ allows the VCC voltage of the amplifier to be set over a wide range.

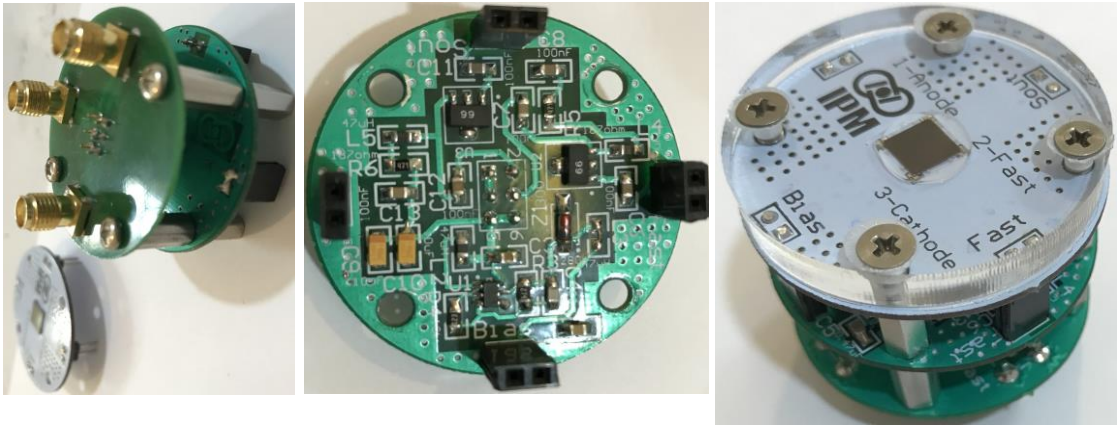
**Figure 10.** Fabricated SiPM detector.

Therefore, voltage 7V of the detector power supply is suitable for this amplifier. After passing through this amplifier, a positive SiPM signal is inverted (negative) at the output with a tenfold amplification [7]. So the output signal from the SiPM is the same as the PMT driver signal. To further optimize the signals, they must be re-amplified. Here the OpAmp is used in the following circuits. Figure 9 shows a schematic of the preamplifier circuits for standard and fast outputs. Figure 10 shows a fabricated SiPM detector.

## 3. Amplifier and Peak Detector

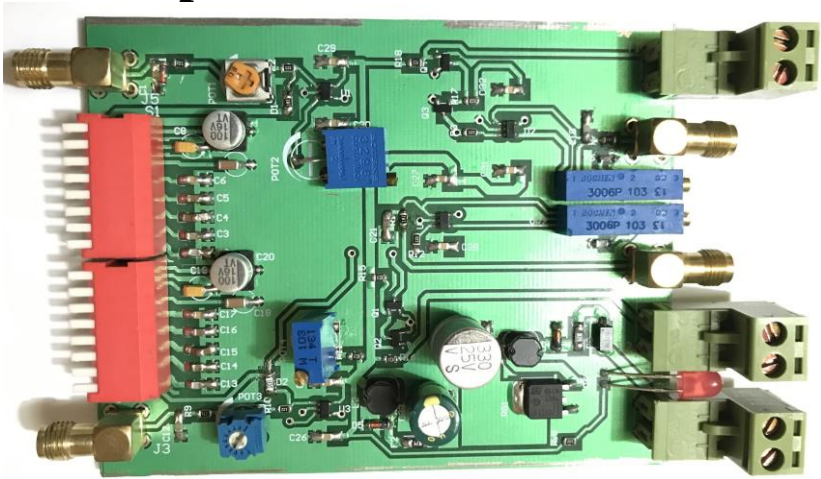
**Figure 11.** Fabricated twin Amplifiers and peak detectors.

The voltage of output signals from PMT and SiPM detectors in particle detection is usually less than 0.5V and, on average, about 100mV. The duration of the signals is in the range of 10 to 100 nanoseconds. Since their height and time are very irregular in particle detection, the conditions are not favorable for counting and measuring the height of the signals. Counters only count digital signals. So the signal must be a square wave at the digital logic level (equivalent to VCC, which is usually 3.2 or 5 volts). For this purpose, the signal must first be amplified as much as possible. However, by amplifying the signals, noises are also amplified. Some OpAmp amplifiers have anti-noise structures and eliminate background noise. [8].

To measure the height of a fast and weak signal, firstly, the measuring device must have a very large input resistance of about 50-100 megohms. Secondly, the measuring device must act very fast to detect the signal peak with a few nanoseconds duration. However, such a system will not be small, simple, and cheap. Since such signals usually occur at a significant time interval relative to the duration of each signal, the signal can be extended during amplification. As a result, by increasing the duration of the signal, the measurement system is given enough time to capture the peak.

Like other parts, amplifiers must have a regulated power supply circuit. The OPA354 amplifier IC requires a power supply of 3.2 volts, and the maximum amplitude of the output signal is not more than 3.2 volts, which is fully compatible with subsequent circuits that process the signal. Figure 12 shows the power supply circuit of the amplifier.

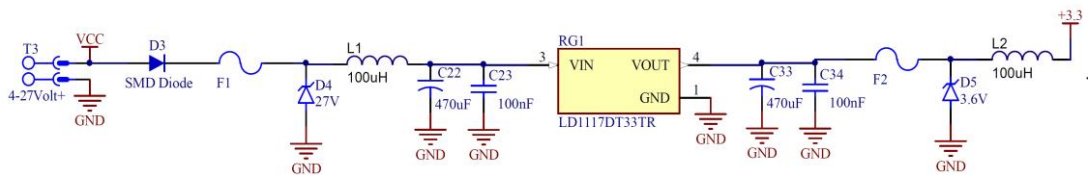

**Figure 12.** Supply voltage for amplifier.

This amplifier device consists of two separate lines. In each line, two amplifier ICs are placed in series, but they are not connected directly to each other, and a circuit is placed between them to increase the signal duration. First, the primary amplification circuit is introduced [8] (Figure 13). This circuit receives a negative signal from the -IN pin, and the amplified output is a positive signal. The effect of POT1 and POT2 potentiometers on the amount of amplification and prevention of over-amplification of background noise is very decisive, and it is necessary to monitor the output of the amplifier and adjust those according to the shape and height of the input signal to get the best output signal from the amplifier.

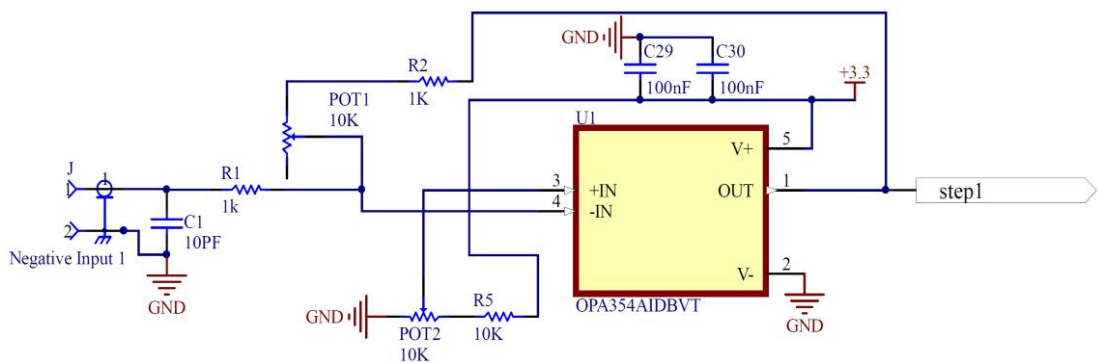

**Figure 13.** First step of amplifier with negative signal input.

The output signal from the first step is passed through a very fast Schottky diode D1 (BAS70-04). This diode's maximum reverse recovery time is about two nanoseconds [9].This time is critical to be able to propagate very short signals as well. However, SS14 with 500 nanosecond reverse recovery time works well for signals longer than 50ns. [10].

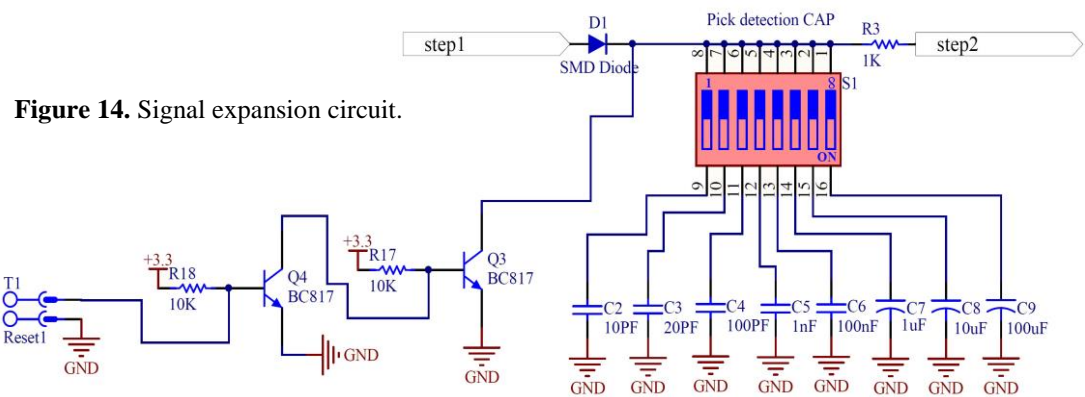

**Figure 14.** Signal expansion circuit.

In addition, this diode, with its potential barrier of about 0.2 V, completely prevents the passage of background noise that is amplified once, and the background remains without those short noises. When the amplified signal passes through diode D1, capacitors C2 to C9 (if connected to the circuit with the piano switch S1) store some electrical charge in proportion to the signal height and time. When the input signal disappears, diode D1 prevents current from returning to the first amplifier. If the diode's recovery speed is not sufficient, small signals will not have a chance of passing successfully and creating an effect on the capacitors.

The circuit on the left, which is finally connected after diode D1, is used to discharge the capacitors. If high-capacity capacitors are activated, they will take a long time to discharge due to the high-resistance connection to stage 2 of the amplifier. This circuit can be used to discharge capacitors manually or by the controller which uses this amplifier. The larger the capacitors in this section, the greater the output signal time (expansion time).

Figure 15 shows the image of the second stage of the amplifier path. The electrical charge stored in the capacitors is discharged slowly due to the input port with a very high resistance in the second stage of the amplifier, and the time expansion depends on the effective capacitance. For short signals of about ten nanoseconds, a capacitor of about 20 Pico Farads can almost double the signal duration. The signal is amplified again at this stage and the effects of background noise are reduced due to the anti-noise function of the amplifier again.

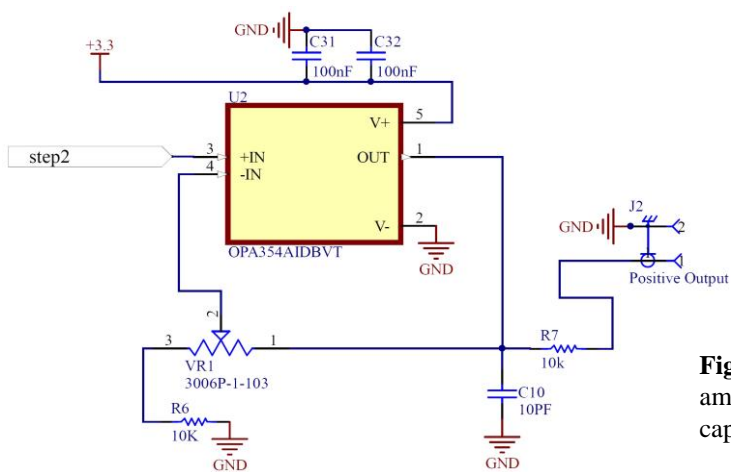

**Figure 15.** Second amplification circuit after capacitors.

To avoid saturation of the amplifier output, the potentiometers of the first and second stages must be properly adjusted so that the height of the output signals is proportional to the height of the input signal. Saturation means the output signal height reaches 3.2 volts (maximum output height).

Signals from cosmic muon collisions with a plastic scintillator BC408 connected to a PMT Hamamatsu R580 (under a potential difference of 1400 V) are shown in Fig. 16.a without any attempt at optimization.

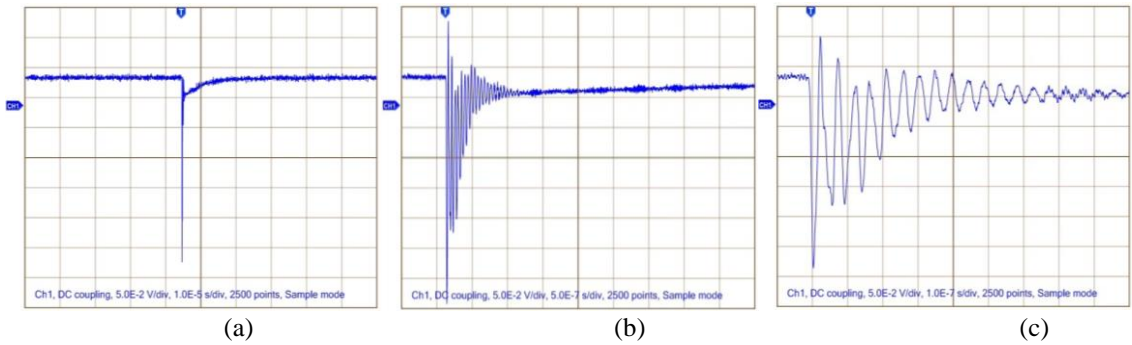

(a)                 (b)               (c)

**Figure 16.** Signals of cosmic muons, passing through plastic scintillator, catch by PMT, in various time resolution.

As shown in Figures 16.b and 16.c, at higher time resolutions, it can be seen that there are rings in the signal tail. These rings can be well removed by connecting a 50-ohm resistor, but this very low resistance causes the signal to be suppressed and the shape and height of the signal to change significantly. With its internal circuits, this designed amplifier can eliminate the ring's effects and does not suppress the signal. The signal amplified and optimized by the amplifier, without increasing the time expansion (disconnection of capacitors C2 to C9), compared to the input signal, is shown in Figure 17.

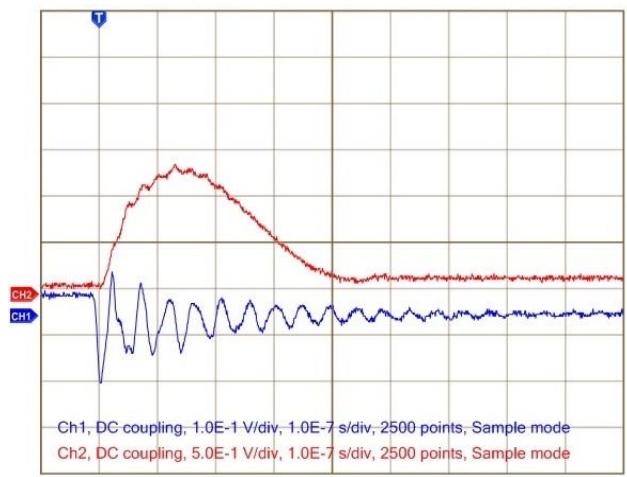

**Figure 17.** Negative signal in CH1 got from PMT and amplified, positive and optimized signal in CH2.

The value of capacitors C2 to C9 should be selected according to the length of the input signal. Capacitors with a capacity of more than 200 pF are not useful for amplifying the signals in Figure 16 because their suppression effect is too significant on fast and weak signals and can also merge them with the subsequent signals due to expansion.

Of course, these values are obtained qualitatively and experimentally because many parameters must be considered for accurate calculation. It is much easier to test ordinary capacitors (available in the electronics market) in this circuit to optimize the capacitance.

Figure 18 shows the effect of adding 130pF (activation of C2, C3, C4) on the output signal compared with the input signal at different time resolutions. As seen in Figure 18, the output signal is much broader than the short input, but not so much that it can be merged with the following signal. In the case of cosmic muons, given their mean flux at sea level [11], we can safely assume that the signals do not merge.

However, if the signal needs to be more or less expanded in a particular application, this is possible by effectively changing the capacitor. However, with this expansion, the signal processing system has a much greater chance of catching peaks.

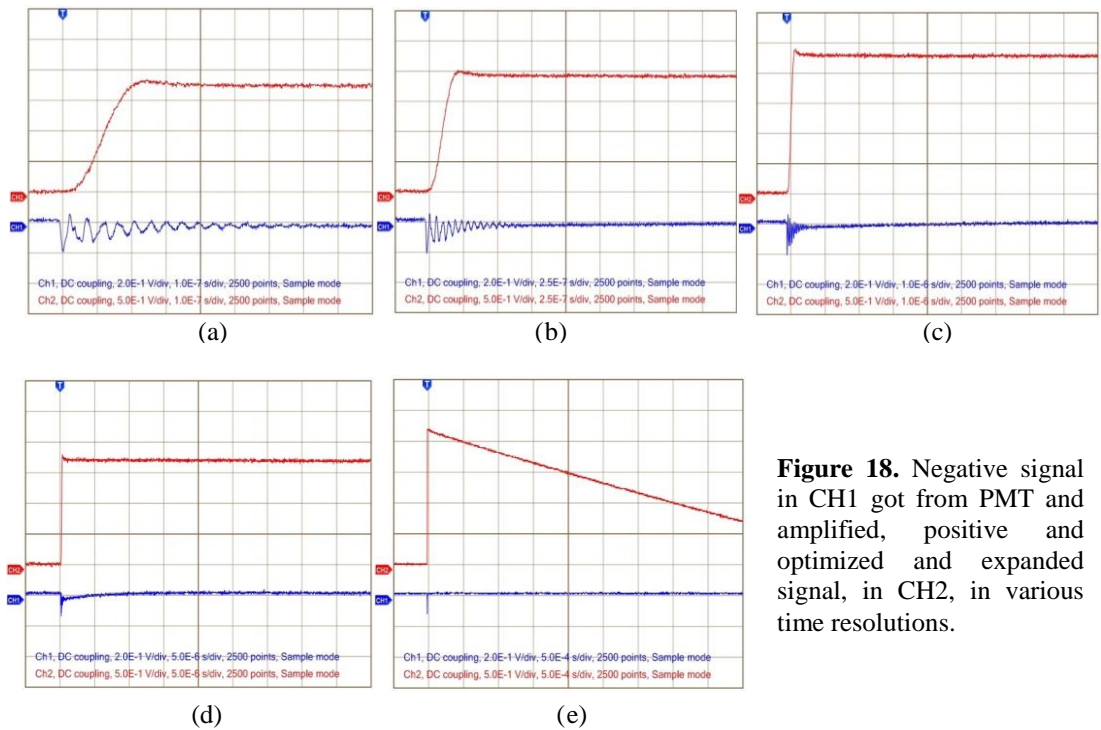

**Figure 18.** Negative signal in CH1 got from PMT and amplified, positive and optimized and expanded signal, in CH2, in various time resolutions.

The choice of the first and second-generation data transmitter ICs operating at frequencies of 100 to 2 GHz is very suitable for amplifying the signals obtained from PMT and SiPM in scientific experiments. Because people have tried for years to optimize these amplifiers to transmit data safely, and fortunately, in terms of frequency, speed, and time performance, they are in the range required for particle detection.

OPA354, for example, is a complex amplifier that, if simplified, could be a schematic of Figure 19 [8]. This complexity in structure is used to optimize signals in scientific experiments. However, the use of simple amplifiers, including transistors and MOSFETs, increases the heavy and challenging task of signal optimization.

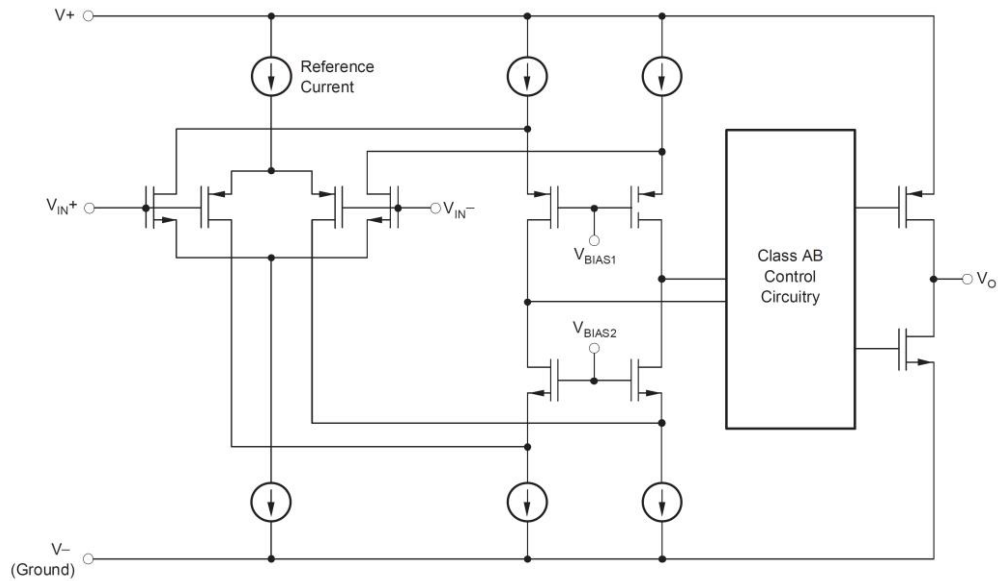

**Figure 19.** Simplifies Schematic of OPA354 internal circuit.

## 4. ADC (Analogue to Digital Converter)

After passing through the peak detector, the signals received from the radioactive sources or cosmic muons on the system mentioned in the previous sections will have a duration of about a few microseconds. If the goal is to measure the signal height (proportional to the particle energy passing through the detector), there is now almost 1 microsecond for peak hunting. In this case, it is appropriate to use an analog-to-digital converter with a sample rate of about 3MSPS (Mega Sample per Second) with a conversion accuracy of 16 bits. This converter is likely to capture most peak signals.

Storing such a large amount of data at such a speed is a challenge. The important thing is the peak of the signal, not its shape, while the signal's shape has been optimized and changed for better results. For initial storage and processing of conversion results, STM32H7x microcontrollers with ADC 16bit 3.2MSPS converter are used. The amount of RAM for fast conversion data storage depends on the programming. In any case, something like half of this memory, i.e., 1 megabyte, will definitely be available. This memory saves about 500,000 16-bit conversions in RAM. This amount of conversions can be done in 1/6 seconds, so for just one conversion and saving data on RAM, it needs a very short time. For every 500,000 conversions, a small interrupt is required to separate the peaks and store them in other memory, such as an external EEPROM (which can be selected 1 MB or more because the number of peaks is not so large and each peak's height occupies 2bytes) [12]. Since the time interval between signals in radioactive sources and cosmic muons is usually much longer than these interrupts in data storage, almost all peaks can be stored using this method.

Faster 16-bit converters are standalone ICs and require a much faster processor to run them. Using them is much more expensive and complicated. Given that the setup mentioned is accurate, there is no justification for using a more expensive and complex system. It should be noted that common digital oscilloscopes can convert even up to 4GSPS and more. Nevertheless, to obtain the peak of each signal, they must receive the entire signal and store its data. Such a process takes up to several seconds for each signal [13]. As a result, about 4 seconds are wasted for each peak, and most of the signals are lost. Therefore much larger time and memory are required to store a sufficient number of signals and study their peak. In addition, the cost of providing such a system is much higher than building a circuit.

## 5. Compare (Make signal to be in logic levels)

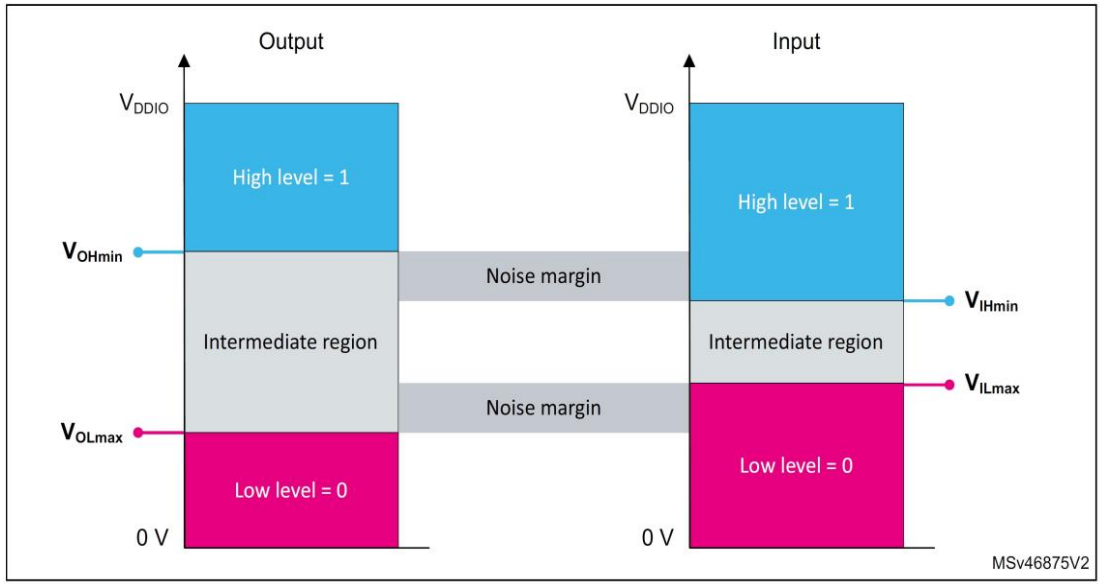

**Figure 20.** Logic level compatibility. For the CMOS technology, the input threshold voltages are relative to $V_{DD}$ as follows: $V_{IHmin} \sim 2/3\ V_{DD}$ and $V_{ILmax} \sim 1/3\ V_{DD}$. For the TTL technology, the levels are fixed and equal to $V_{IHmin} = 2V$ and $V_{ILmax} = 0.8V$.

Counting the number of signals is sometimes the main purpose of using a detector. For this purpose, signals that may be strong, medium, or very weak (not far from the background noise) should be separated from the noise and amplified in such a way that they become a square wave at the height of the digital signal.

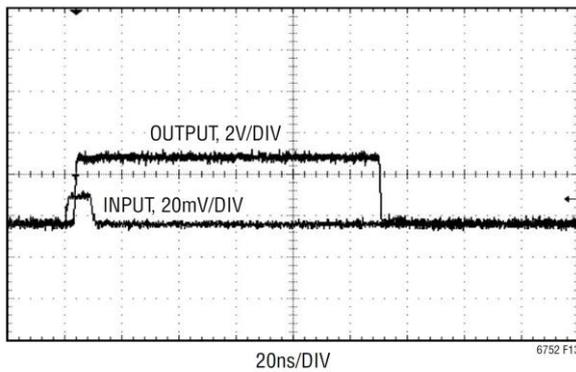

**Figure 21.** Comparison and expansion of signal with LTC6752 by the figure 22 circuit.

The concept of the digital signal is not simple. Digital is not just 0 and 1 in principle. As shown in Figure 20 [14][12], the difference between 0V and VDD (generally 3.2V) is divided into four parts [12]:

- the upper level (equivalent to 1 digital)

- the lower level (equivalent to 0)

- the central level (signals of this level are ignored)

- the area of Noises (this area causes errors because sometimes it may be considered as one or zero and sometimes ignored).

Before transmitting the signal to the counting circuits, we need to ensure that the signal is in the Logic area, i.e., areas 1 or 0. The comparator circuit is used for this purpose.

The comparator IC compares the input signal with its reference voltage in the circuits. Then it converts the signals above this voltage to logic level 1 and takes the rest of the signal to level 0. This way, the input signal will switch to a squared signal at the output. Ideally, the LTC6752 IC can raise the 20 mV signal to a logic level. Similar to what we did with the amplifier, someone can use two comparator ICs to expand the signal duration [15]. Figure 21 [15] is an example of output signal of the circuit in Figure 22 [15].

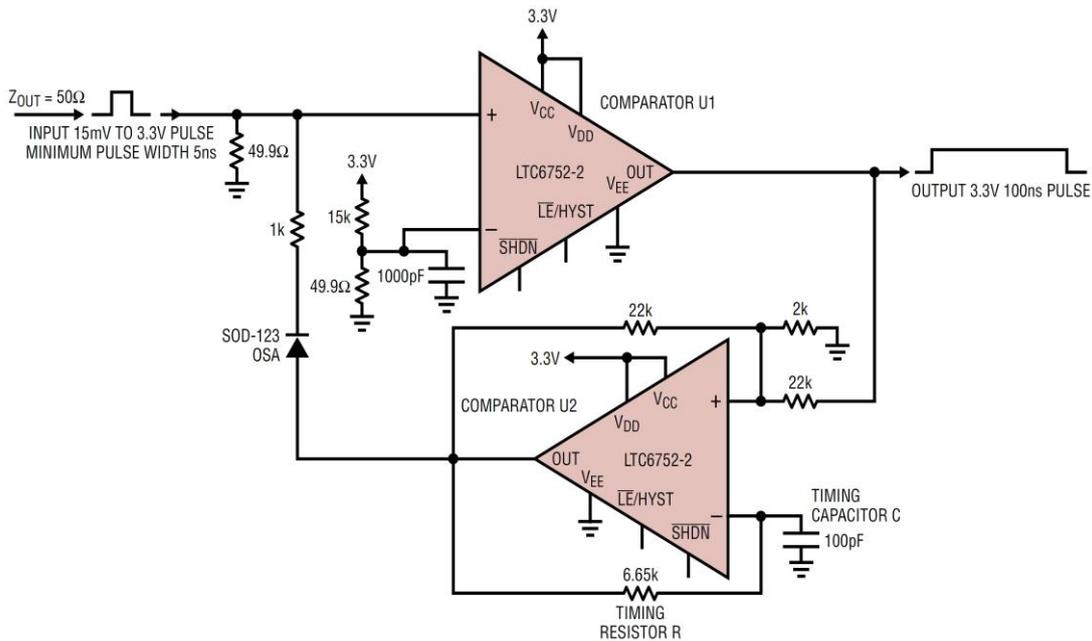

**Figure 22.** Circuit of comparison and expansion of signal with LTC6752.

The time expansion of the signal is done better and easier with the method presented in section 3. In the OpAmp circuit, there was no need for strict conditions to eliminate noises, and the ideal shape of the input signal was not necessary, but working with comparators to amplify and expand the duration of the signal is more complicated. The internal circuits of comparator ICs do not successfully eliminate background noises. However, after passing the signal through the introduced amplifier, it is necessary to pass it through a one-step circuit of the comparator IC to convert to the logic level.

Figure 23 shows a schematic of the comparator circuit used here. The input signal to this circuit is positive. The critical points in this circuit are:
- Firstly, a potentiometer (R46) can adjust the negative pole of the signal (IN-) from 0 to 3.2 volts. This helps avoid signals lower than (IN-) and is very useful for noise cancellation. This method, like a pair of scissors, shortens the height of all signals, which is very good for removing background noise.
- Secondly, the signal at the output returns to the input as feedback by passing through another potentiometer (R44). Adjusting this potentiometer determines the amount of feedback and, consequently, the amount of amplification. This way, the optimal output signal can be achieved by adjusting two potentiometers, which affect the input signal.

With the help of the circuits presented so far, it is possible to obtain completely optimal signals for energy measurement and their counting. The comparator IC is fast enough (Low Propagation Delay, equivalent to 2.9 nanoseconds) to successfully pass signals with a duration of 10 nanoseconds.

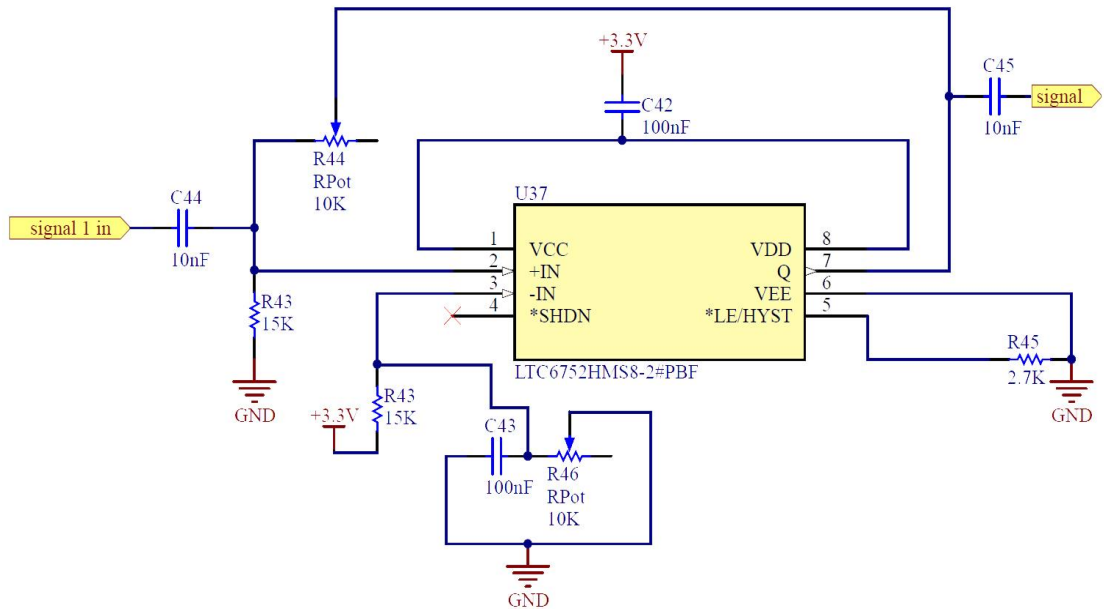

**Figure 23.** Circuit of comparison of signal with input control by LTC6752.

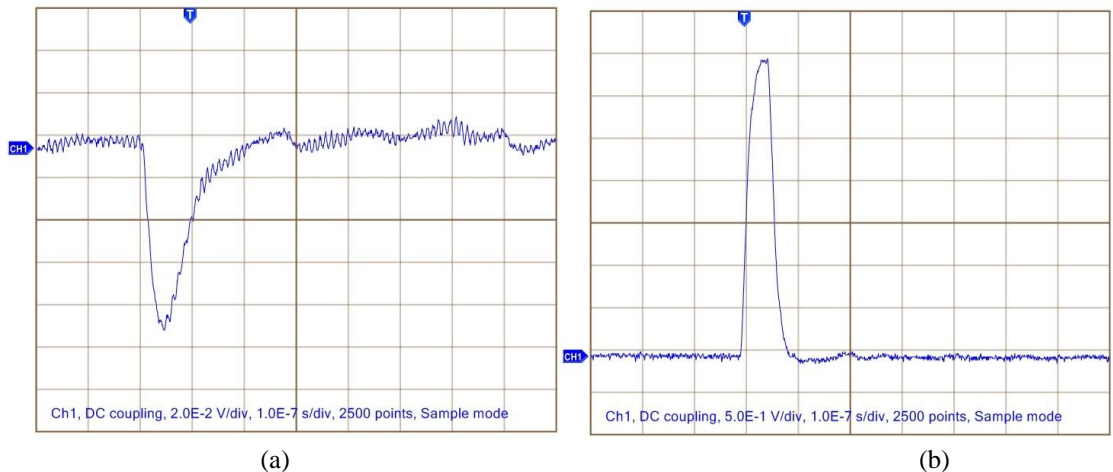

(a)                                  (b)

**Figure 24. a)** Signal take from the detector. **b)** Signal come out of the amplifier and comparator.

This length of time corresponds to the shortest signals received from such detectors. Figure 24 shows the signal received from the detector (before entering the amplifier) and the signal after passing through the amplifier (without time expansion) and the comparator.

## 6. Synchronicity & AND (Logic Gate in High speed in response)

When two or more detectors are used to obtain the particle trajectory, their signals must be approximately synchronous. Suppose the structure of the detectors and the connection cables are the same. In that case, the time interval between the obtained signals can indicate the time interval the particle has moved between the detectors. This measured time can be used to calculate particle energy.

Particles such as cosmic muons with an average energy of 4GeV travel 1 meter in less than 20 nanoseconds. This issue is practically examined. For any particle with enough energy to pass through two or more detectors, the velocity of the particle is high enough that it travels the distance between the detectors in the nanoseconds range. Therefore, to measure the synchronicity of the signals, a maximum speed of conversion to the logical level must be achieved.

It is better to convert the signals to digital to measure the time interval of signals. The height of the signals is not significant here. Even particles with the same energy may emit signals of different amplitudes in the detectors due to the fluctuation of the energy loss of particles. It is much easier for electronic circuits to detect the synchronicity of digital signals. However, using chips containing double amplifiers or compares in experiments disrupts the measurement of time intervals. For example, when we used the chip of dual Comparator IC LT1715 for two signals, it eliminated the time interval between them, which was about 15 nanoseconds. Nevertheless, when we used two chips, the time interval was well seen, and the time intervals of almost simultaneous signals were obtained depending on the energy of the incident particles.

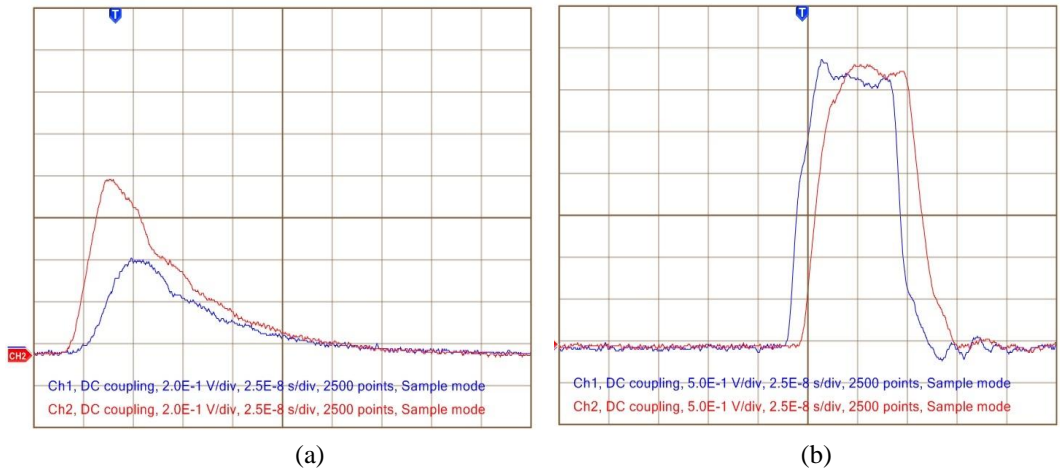

(a)          (b)

**Figure 25. a)** Synchronal signals take from the two series detectors. **b)** Digitized synchronal signals take from the two series detectors.

We should consider the propagation delay time in the comparator and the speed of the amplifier. The amplifier has been discussed before, but LTC6752 is a good choice for the Compare, which has a 2.9ns propagation delay time [15]. If the signals are digitized, it is enough to compare them in an AND gate to detect their synchronicity. Propagation delay time should be as short as possible. For working with two input signals, ONSemi NC7SP08 Tiny Logic 2-Input AND Gate with 4ns propagation delay time is suitable [16]. If both input signals are in the range of logic level 1 and their time overlaps, the output signal duration is approximately the overlap time of the two input signals. The output of this component is a signal similar to Figure 24.b.

A high-speed digital oscilloscope (at least 100 MHz) can measure the time interval between two overlapped signals. Circuits containing high-speed flip-flops can also be used. We are still investigating this circuit. Figure 25.a is an example of the simultaneous signals of two detectors. The time difference between them can also be seen in the figure. Figure 25.b shows two other signals that have been converted to digital. Figure 26 shows a schematic of the AND circuit just after two comparators.

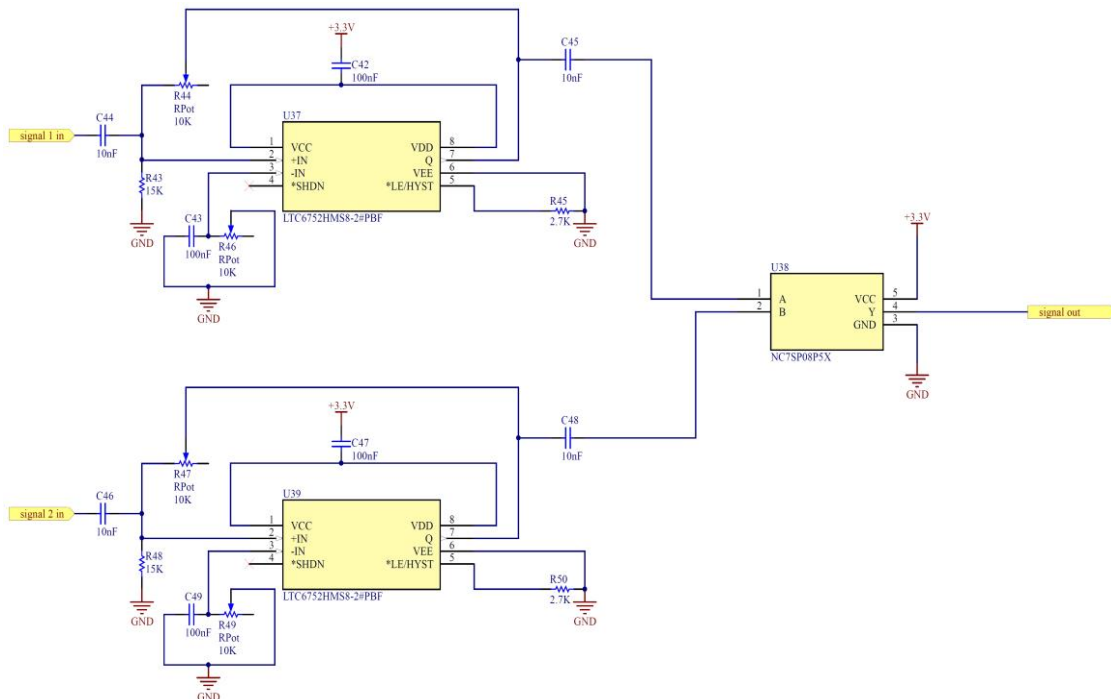

**Figure 26.** AND circuit just after the comparators.

## 7. Fast Counter and Timer, with Flip-Flop

Internal counters of microprocessors and FPGA can be used to count digitized signals that have passed the synchronization stage or digitized signals directly from a single detector. However, according to the studies, a fast, economical and reliable counter can be built Based on a series of Flip-Flop. The 74 LVC1G74 flip-flop with 2.5 ns propagation delay is suitable for counting signals with a duration of at least eight nanoseconds [17]. Figure 27 shows a schematic of this counter with a series of 31 flip-flops.

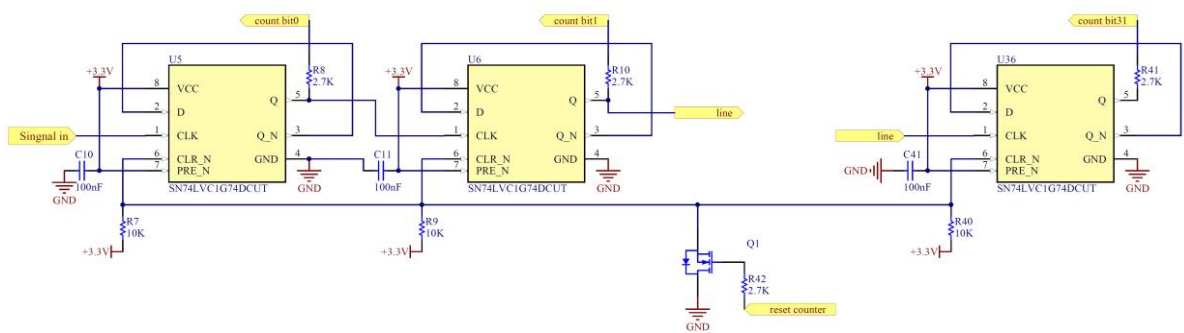

**Figure 27.** High speed counter with series of flip-flip 74LVC1G74.

For each count, all flip-flops have to be reset by MOSFET Q1. The input signal of any frequency is divided in half during each step. Only one signal goes to the next stage of Flip-Flop for two input signals. It creates a binary value in the output of the flip-flop series. In this way, the processor, whose 31 pins are connected to countbit1-31 (according to figure 27), assigns a value of $2^n$ to each count-bit (n). The processor independently tracks the count at any time and can read the pins without interrupting the counting process.

A signal generator tested this counter, and up to the maximum frequency of that device, equivalent to 100MHz, counting resulted in great accuracy in different time intervals. One can create a similar circuit and send pulses to it (signal in) with a precise oscillator to measure time. The frequency of this pulse can determine the time value of each count-bit. The processor only needs to check the value of these pins connected to the count-bits (0 or 1) and then obtains and reports the time elapsed at any time, with a simple calculation.

If a 1MHz oscillator is used, the time measurement accuracy will be up to one microsecond, and 32 series flip-flops can measure time up to 1000 seconds. Of course, using the microcontroller's built-in timers is possible, but its timing accuracy may not be acceptable for very accurate tests.

## 8. Conclusion

Based on the circuits provided, we built a programmable analyzer for comparison, AND, Count, ADC, and other related tasks. Figure 28 is a picture of the main circuit of this device. Here is an application of the detector, amplifier, and analyzer we made: Bulk material imaging, especially thick material, more than one meter of rocks and soil, a few centimeters of metal, and more.

To test whether the detectors and circuits are accurate enough, we first examined the effect of thinner layers on the cosmic muon flux. We considered Layers with a thickness of 1, 5, and 9 cm of lead in our experiments. Figure 29 is the experimental setup, and Figure 30 is the result of the cosmic muon flux on the Earth's surface that has passed through the layers. The number of muons passing through the material per hour was counted for different thicknesses of lead (for each thickness, the result was averaged ten times, and a standard error was obtained).

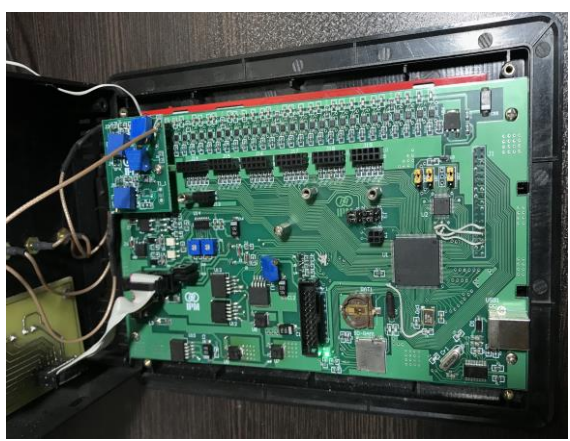

**Figure 28.** The analyzer built with the circuits

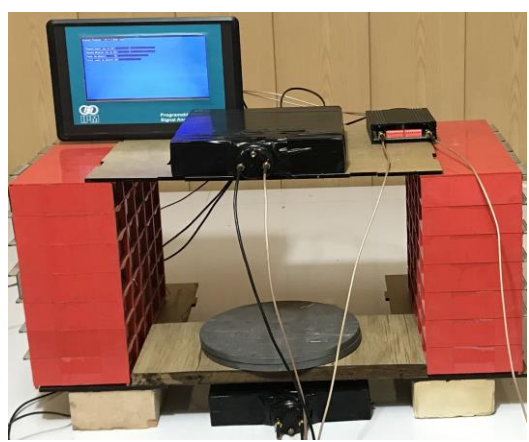

**Figure 29.** Setup of the experiment that measures the effect of Material thickness on the cosmic muon flux.

The diagram shows that after 10 minutes of counting, the error bars are almost separated, and it can be confirmed whether the matter's thickness is 1, 5, or 9 cm thick.

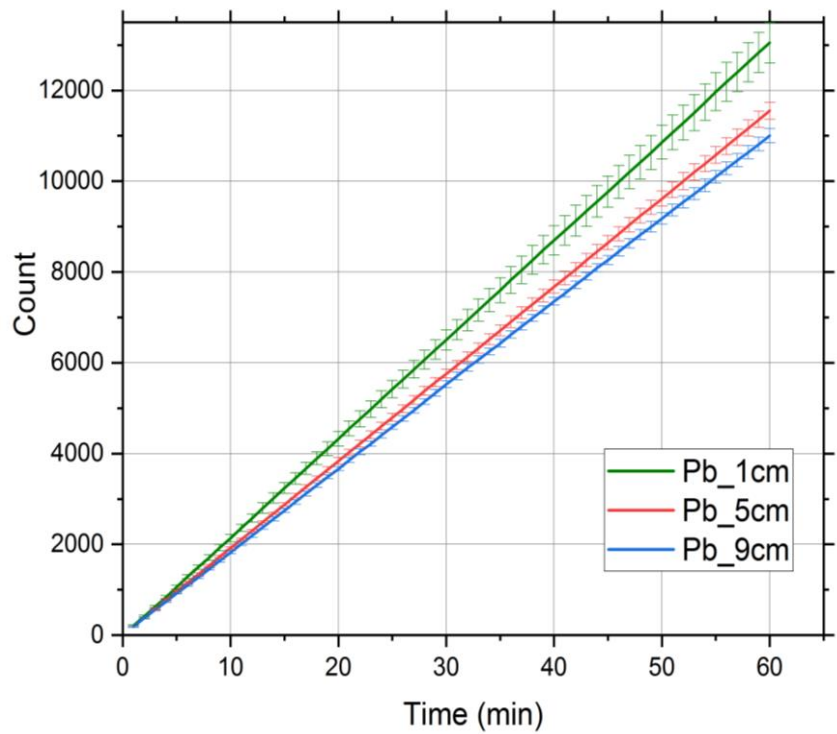

**Figure 30.** The result of the experiment that measures the effect of Material thickness on the cosmic muon flux for Pb 1,5and 3cm.

In another experiment, we measured the relative amount of matter at the top of a tunnel using the setup in figure29. The length of the tunnel is about 500 meters. The muons flux was counted for 10 minutes at intervals of 20 meters. Figure 33 shows the result of this measurement. This diagram is drawn logarithmically, corresponding to the mountain's dimensions above the tunnel. We used a logarithmic scale for the vertical axis on this diagram to get a better fit because the reduction of muon flux in rocks is not a linear function of the depth. Figures 31 and 32 are topographical maps of the tunnel. Figure 34 is a graph derived from the approximate surface height from Figure 33

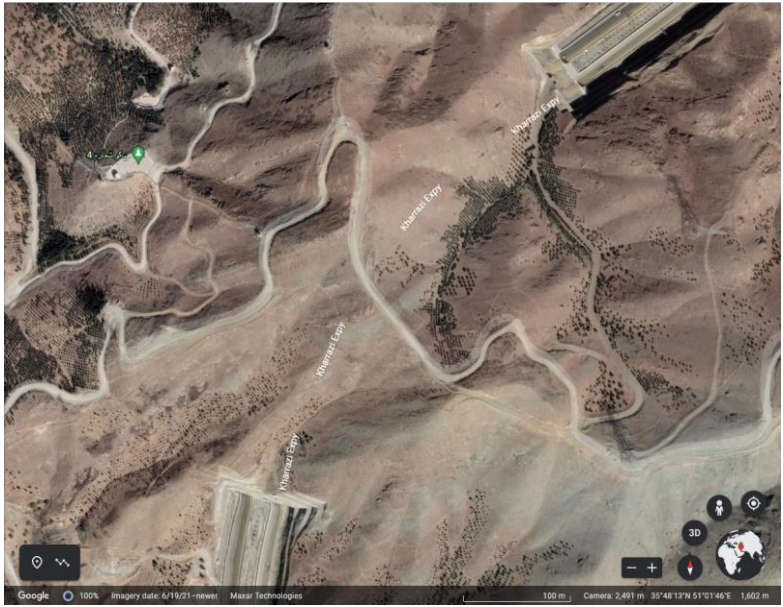

**Figure 31.** Google earth 3D photo of the tunnel.

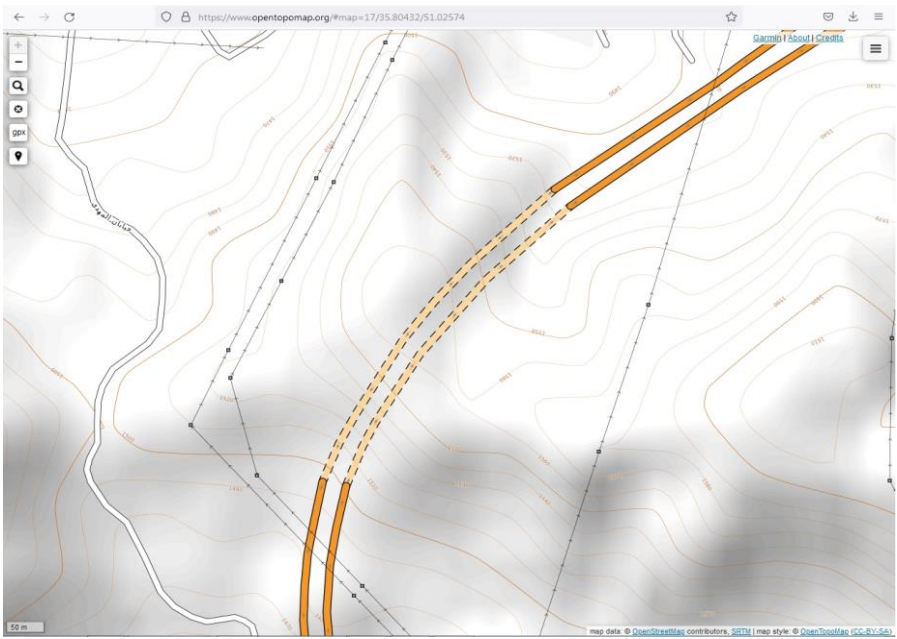

**Figure 32.** Opentopomap tophraphical photo of the tunnel.

The topographic map is a bit inaccurate, but the procedure for height variation is somewhat consistent with the diagram obtained from the experiment. However, the results of these experiments were expressed only for the purpose of testing the correct operation of the circuits provided in this text.

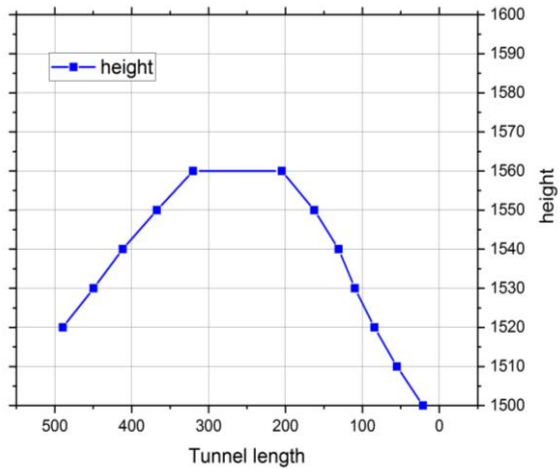
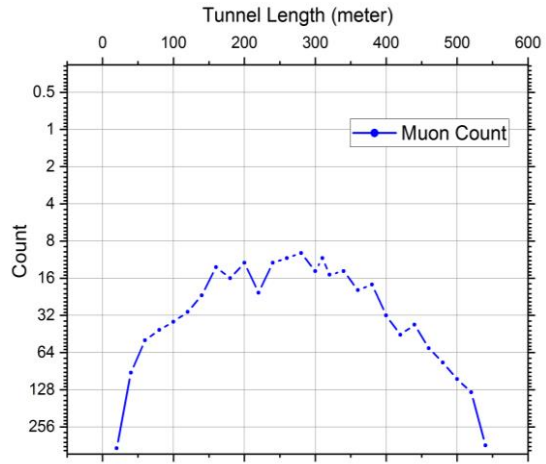

**Figure 34.** The curve that took from topographical map in figure 32.

**Figure 33.** The curve that predicted from cosmic muon imaging from the mountain in the top of the tunnel.

The most important achievement here is to convert the standard signal taken from the detectors into an almost ideal optimized signal for:
- Signal counting; Signals similar to those shown in Figure 24.a are optimized to signals like Figure 24.b.
- Peak detection for energy measurement; optimization like as shown in figure 18.
- Time of flight for particles like those shown in Figure 25.

These optimizations in detector readout will allow us to obtain correct results from future experiments by such instruments.

In the continuation of this research, it is appropriate to try flip-flop-based circuits with the aim of measuring the energy of single particles. For this purpose, digitized signals can be used before crossing the AND. They have a time difference proportional to the incident particle's energy.

## 9. Acknowledgment:

We are especially grateful to Dr. Mohammadi Najafabadi from the School of Particles and Accelerators at IPM for the continual interest shown in the project, and the financial support he provided.